\documentclass[12pt]{article}
\usepackage{a4wide,graphicx}

\newcommand{\mycaption}[1]{\caption{\small #1}}
\newcommand{\GeV}{\ensuremath{\mathrm{\;GeV}}}
\newcommand{\TeV}{\ensuremath{\mathrm{\;TeV}}}
\newcommand{\mch}[1]{\ensuremath{m_{\tilde\chi_#1^+}}}

\newcommand{\beq}{\begin{equation}}
\newcommand{\eeq}{\end{equation}}
\newcommand{\nn}{\nonumber}

\newcommand{\snu}{\tilde{\nu}_e}
\newcommand{\chplus}{\tilde{\chi}^+}
\newcommand{\chminus}{\tilde{\chi}^-}
\newcommand{\epm}{e^+e^-}
\newcommand{\afb}{A_{\rm FB}}
\newcommand{\alr}{A_{\rm LR}}

\begin{document}
\setlength{\unitlength}{1mm}


\begin{titlepage}
\begin{flushright}
{\bf KA--TP--20--2000\\
hep-ph/0011092}
\end{flushright}
\vspace{2cm}
\begin{center}
{\Large \bf Radiative Corrections to Chargino Production 
  \\[0.4cm]
in Electron-Positron Collisions} \\[1.5cm]
{\large T.~Blank\footnote{E-mail: Torsten.Blank@physik.uni-karlsruhe.de},
W.~Hollik\footnote{E-mail: Wolfgang.Hollik@physik.uni-karlsruhe.de}
}  \\[1.0cm]
{\normalsize\em 
        Institut f\"ur Theoretische Physik, Universit\"at Karlsruhe,\\
        D-76128 Karlsruhe, Germany}
\end{center}
\vspace{2cm}
\begin{abstract}  
The results of a complete one-loop calculation for 
the production of chargino pairs 
in electron-positron collisions, including the option of polarized beams,
are presented. The calculation has been performed in the
on-shell renormalization scheme.
Applications to the integrated
cross section, to the forward-backward asymmetry
for unpolarized beams, 
and to the left-right asymmetry show that 
the higher-order effects have a sizeable influence on the
theoretical predictions and therefore should be properly taken
into account for detailed studies in the MSSM.
\end{abstract}
\end{titlepage}


\section{Introduction}
In the MSSM, charginos are the fermionic
superpartners of the $W^\pm$ bosons and the charged-Higgs bosons,
in specific superpositions forming the mass eigenstates.
High-energy electron-positron colliders are the best environment
to make a precise determination of their properties.
In $\epm$ annihilation, charginos
are produced in pairs, in lowest order described
by $\gamma,Z$ $s$-channel exchange and $\snu$ $t$-channel exchange.
In that approximation, the cross section is determined by the gauge
couplings and by the masses of the charginos and of the $e$-sneutrino.
From precise measurements of the masses,
production cross sections and
asymmetries, the fundamental parameters of the model can be 
reconstructed \cite{reconstruct}.   
In view of these prospects it is necessary to include the radiative 
corrections in order to get theoretical predictions matching the 
experimental accuracy.

It has been shown in previous studies that the lowest-order 
predictions can significantly be influenced by higher-order
contributions. These studies concentrate on the one-loop  
diagrams with virtual fermions and sfermions  \cite{oneloop},
which can involve large Yukawa couplings.
In this note we present the result of a complete
one-loop calculation, where all MSSM particles 
with electroweak couplings are included in the
virtual states. As a calculational frame, the on-shell 
renormalization scheme has been applied where all particle
masses are physical on-shell quantities.  
Cross sections and asymmetries are thus directly
related to the physical masses of the charginos and of the other
particles entering the loops.  
We briefly outline the theo\-retical basis of the calculation and 
discuss the numerical size of the loop effects in the 
integrated cross sections (for unpolarized beams)
and for the angular distributions in terms of the
forward-backward asymmetry, as well as for the left-right asymmetry
in case of longitudinal beam polarization.

\section{Amplitudes and cross sections}
The differential cross section for 
$\epm \rightarrow \chplus_j \chminus_i$ ($i,j =$ 1 or 2) 
with given $e^\pm$ helicities $\eta_\pm$ 
can be written in terms of the helicity amplitudes ${\cal M}$
as follows,
\beq
 \frac{{\rm d}\sigma}{{\rm d}\Omega}(\eta_+,\eta_-) 
  = \frac
 {\lambda(s,m_i^2,m_j^2)}{64\pi^2 s^2} \, \sum_{s_+,s_-} \,
 | {\cal M}(\eta_+,\eta_-,s_+,s_-) |^2 \, ,
\eeq
with 
$\lambda(s,m_i^2,m_j^2) = 
 (s^2+m_i^4+m_j^4-2sm_i^2-2sm_j^2-4m_i^2m_j^2)^{1/2}$.
The sum extends over the final-state helicities $s_\pm$.
Unpolarized $e^\pm$ in the initial state
require a further summation over $\eta_\pm$, together with an
additional factor 1/4.
Of specific interest are the integrated cross section
\beq
\label{xsec}
 \sigma = \int {\rm d}\Omega\, \frac{{\rm d}\sigma}{{\rm d}\Omega} ,
\eeq
the forward-backward asymmetry
\beq
\label{afb}
 A_{\rm FB} = \frac{\sigma_{\rm F} - \sigma_{\rm B}}
                   {\sigma_{\rm F}+ \sigma_{\rm B}} \, ,
\eeq
with the cross sections $\sigma_{\rm F(B)}$ integrated over
the forward (F) and backward (B) hemispheres in the
center-of-mass system, 
and the left-right asymmetry
\beq
\label{alr}
 A_{\rm LR} = \frac{\sigma_{\rm L} - \sigma_{\rm R}}
                   {\sigma_{\rm L}+ \sigma_{\rm R}} ,
\eeq 
with the integrated cross sections $\sigma_{\rm L(R)}$
for left- and right-handed electrons.

\medskip
At lowest order, 
the amplitude for chargino-pair production can be described by
$s$-channel photon and $Z$-boson exchange and by $t$-channel
exchange of a scalar neutrino ($\snu$), as displayed in the 
following  Feynman diagrams:
\begin{center}
  \raisebox{9mm}{\includegraphics[width=6.5cm]{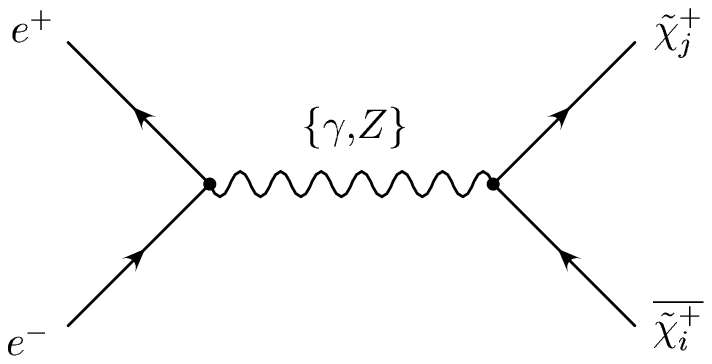}}
  \hspace{1cm}
  \includegraphics[width=4.5cm]{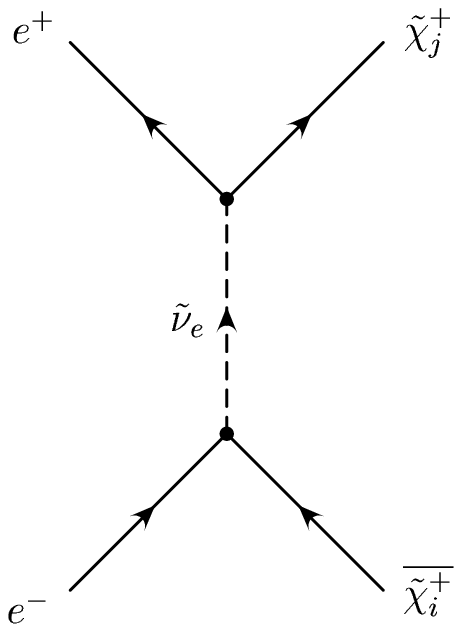}
\end{center}
For a heavy $\snu$, the $t$-channel-exchange contribution
decouples.

\medskip
At the one-loop level, the full particle spectrum of the MSSM enters
into the amplitudes, 
and the process becomes dependent on all the parameters of the 
model. Schematically, the loop contributions are depicted below,
grouped into three classes.

\begin{itemize}
\item Self-energy contributions:\hfill\\
  \begin{center}
    \raisebox{1.3cm}{\includegraphics[width=7cm]{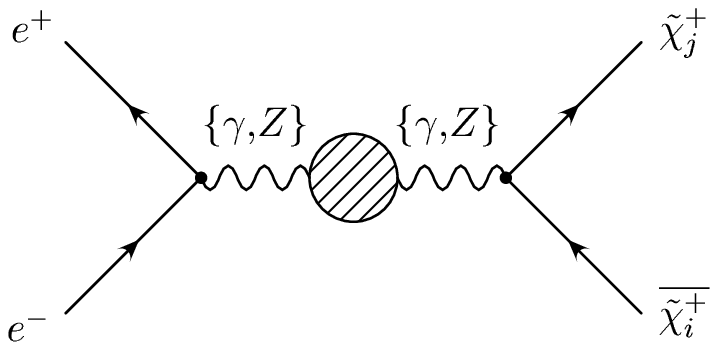}}
    \hspace{1.0cm}
    \includegraphics[height=6cm]{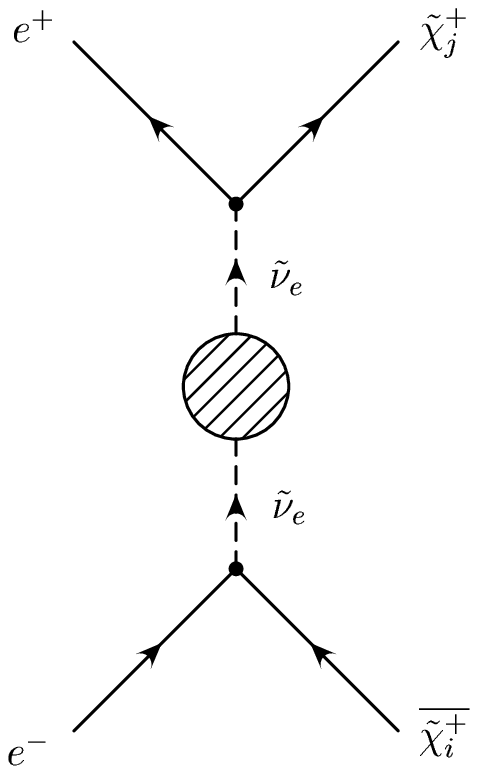}
  \end{center}
\item Vertex corrections:\hfill\\
  \begin{center}
    \includegraphics[width=7cm]{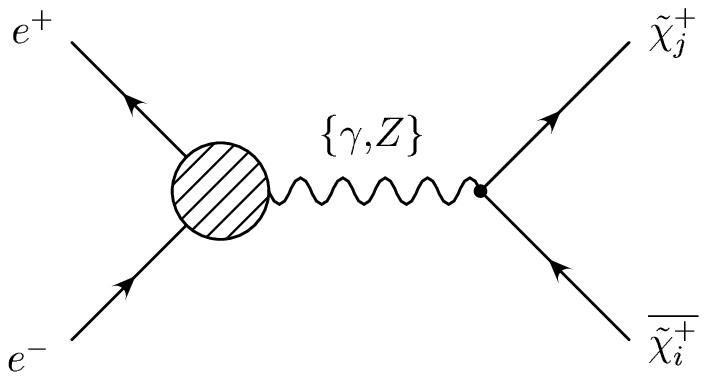}
    \includegraphics[width=7cm]{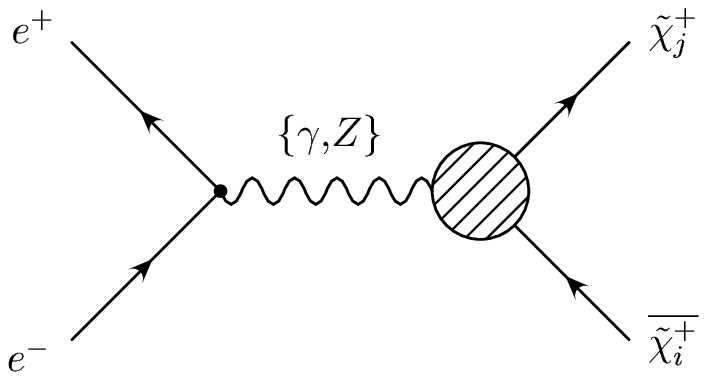}\hfill\\[3mm]
    \includegraphics[height=6cm]{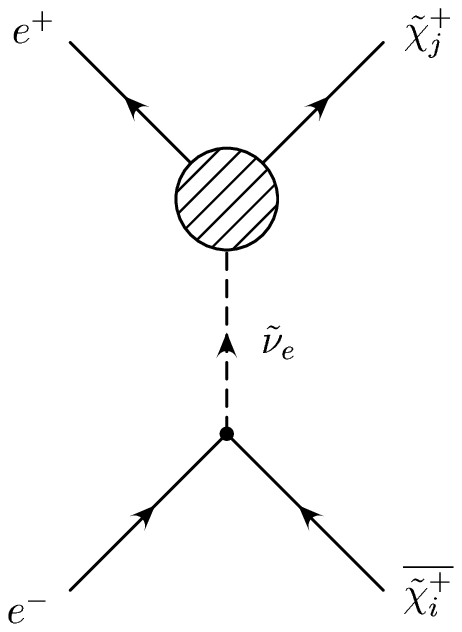}
    \hspace{2.2cm}
    \includegraphics[height=6cm]{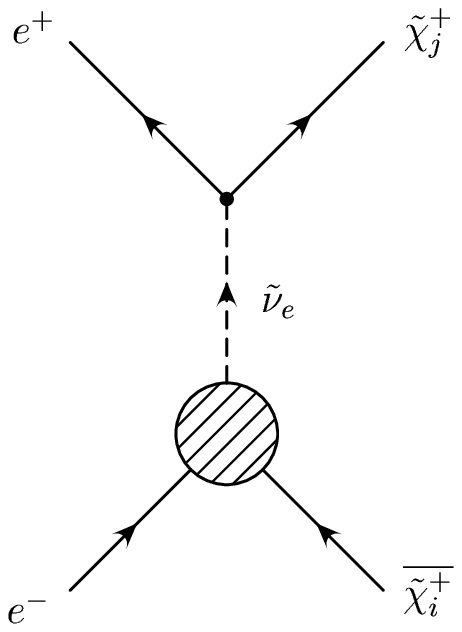}
  \end{center}
The bubbles summarize the one-particle irreducible 3-point functions
including their counterterms,
together with the wave-function renormalization of the external lines
and chargino mixing.
\item Box-diagram contributions:\hfill\\
  \begin{center}
    \includegraphics[height=3cm]{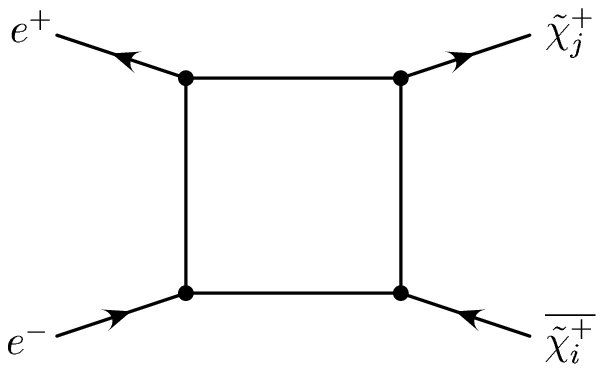}
    \includegraphics[height=3cm]{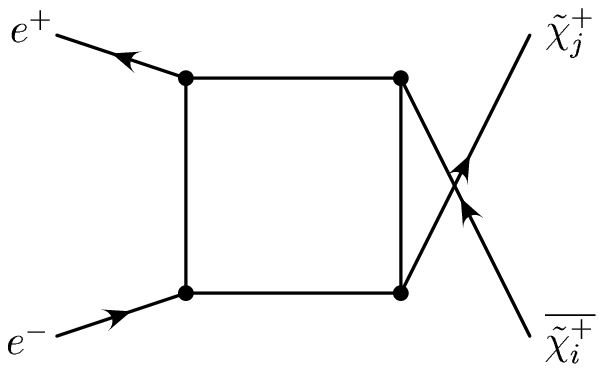}
    \includegraphics[height=3cm]{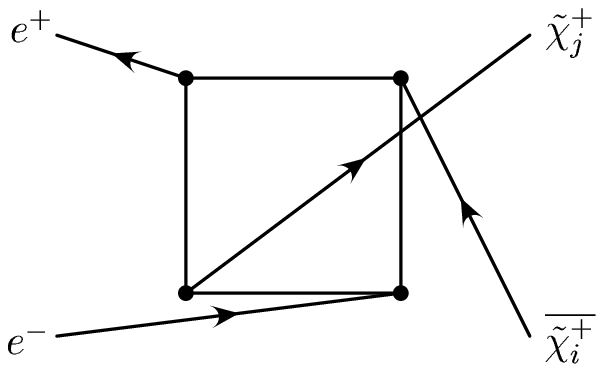}
  \end{center}
For illustration, the  particle insertions 
for the first box topology are depicted
(with gauge bosons $V$, Higgs bosons $H$,
charginos/neutralinos $\chi$,
leptons $l$, and sleptons $\tilde{l}$):           \hfill\\
\end{itemize}
\begin{center}
  \includegraphics[height=4cm]{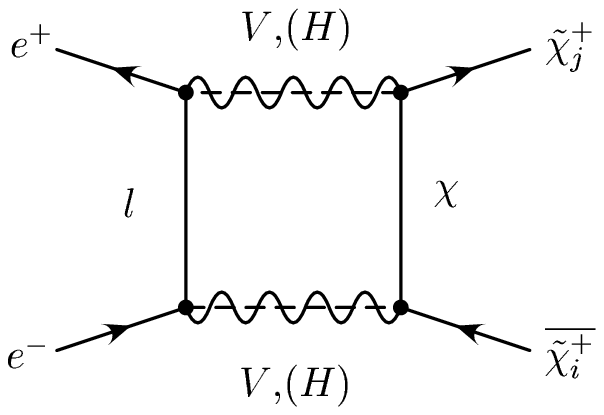}
  \hspace{5mm}
  \includegraphics[height=4cm]{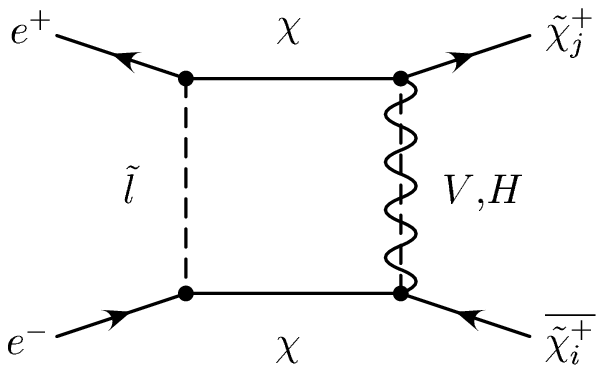}\hfill\\[3mm]
  \includegraphics[height=4cm]{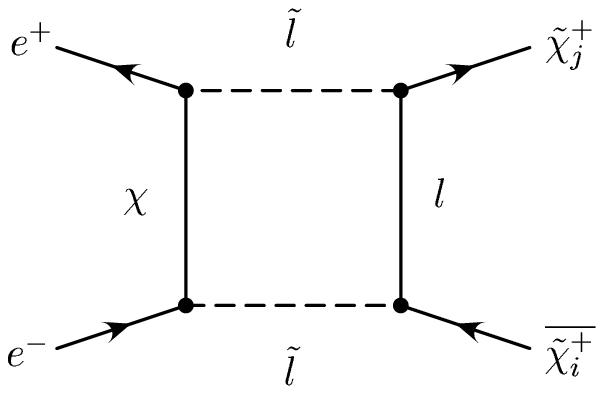}
  \hspace{5mm}
  \includegraphics[height=4cm]{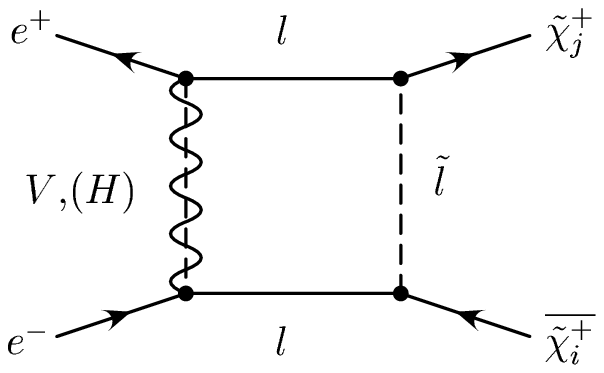}
\end{center}
The diagrams with a Higgs line attached to an electron line are
negligible, owing to the small electron Yukawa coupling.

The previously calculated fermion/sfermion-loop terms
are contained in the self-energy diagrams, in the $V\chi\chi$
vertex corrections, and in the
wave-function renormalization from the external-chargino
self-energies.

\medskip
For a complete  one-loop calculation, 
the full particle spectrum of the MSSM has to be taken into account
in the loop diagrams, 
and the process becomes dependent on all the parameters of the 
model.  The parameters are chosen in the following way:
\[
\begin{array}{ll}
  {\mbox{gauge sector:}} & \alpha, M_Z, M_W  \nn \\
  \mbox{Higgs sector:} & M_A, \tan\beta \nn \\
  \mbox{gaugino sector:} & M_2, M_1, \mu \nn \\
  \mbox{sfermion sector:} & M_{\tilde{L}},M_{\tilde{Q}}, A_f \, . \nn
\end{array}
\]
The sfermion-mass parameters for the slepton and squark generations
$M_{\tilde{L},\tilde{Q}}$, for simplicity, are chosen as equal in the
numerical analysis.

Applying the on-shell renormalization scheme to the charginos,
the chargino masses $m_{\chplus_1}$,$m_{\chplus_2}$ are considered
to be the pole masses, which fixes the mass counterterms by the on-shell 
values of the chargino self-energies. 
In this way, the parameters $M_2$ and $\mu$
are determined in terms of the chargino masses, in the same way as in the
lowest order relations. Their physical meaning, however, is different
from e.g.\  those in a $\overline{MS}$ renormalization scheme, which
becomes important when comparisons between different schemes 
are performed. In a similar way, the $\snu$ on-shell
mass renormalization relates  the slepton mass parameter
to the $\snu$ mass, 
$m_{\snu}^2 = M_{\tilde{L}}^2+\cos(2\beta) \, M_Z^2/2$.
The other parameters enter only at the one-loop level and hence do
not need renormalization. 

\medskip 
A special subtlety arises from the contributions with virtual photons
attached to two external charged particles, which are infrared divergent.
To obtain a infrared-finite result, real-photon bremsstrahlung
from the external legs has to be taken into account. After integrating over 
the photon phase space, the sum of the one-loop and the bremsstrahlung
cross section is IR-finite.
Different from standard-fermion production, these photonic contributions
cannot be removed and treated separetely as ``QED corrections'',
because the one-loop result without the virtual photons would not
be UV-finite even after renormalization. The presence of the photon 
is required to cancel
the divergence owing to the photino component of the virtual neutralinos.
This is a new aspect, which does not appear in the approximate
calculation with (s)fermion-loops only.

\section{Numerical Results}
In this section we provide an illustration of the higher-order effects
by choosing specific examples of parameters.
We consider the unpolarized integrated cross section (\ref{xsec})
and forward-backward asymmetry (\ref{afb}), 
and the left-right asymmetry (\ref{alr}).
As a fixed input parameter, the mass of the pseudoscalar Higgs boson
is set to $M_A=150$ GeV.
The mass parameters $M_{\tilde{L}}=M_{\tilde{Q}}$, assumed to be
equal for all generations and for left/right chiralities,
are denoted by the common mass scale $M_{\rm SUSY}$; 
sfermions are treated as unmixed.
Moreover, between the gaugino masses
$M_1$ and $M_2$, the usual GUT relation is assumed. 

\medskip \noindent
(i) {\it Integrated cross section:}
For demonstration of the loop effects, we define the relative
correction 
\beq
\label{delta}
 \Delta = \frac{\sigma - \sigma_0}{\sigma_0} 
\eeq
to the lowest-order cross section $\sigma_0$. Thereby, $\sigma_0$
is an improved Born cross section, which already incorporates
the large effects arising from the light standard fermions, i.e.\
the following specifications are made:
\begin{itemize}
\item
The fine-structure constant $\alpha$ is replaced by the QED-running
effective fine-structure constant $\alpha(s)$ in the photon-exchange
diagram;
\item
The normalization of the weak coupling to the  $Z$ boson is chosen in terms
of the Fermi constant: $ (G_F M_Z^2)^{1/2}$;
\item
The normalization of the $e\snu \chminus$ coupling is chosen as
$ (G_F M_W^2)^{1/2}$.
\end{itemize}
Since, for a IR-finite result, we have to include photon bremsstrahlung,
which is treated in the soft-photon approximation
(with a cut $k_\gamma < 0.05 \sqrt{s}\;$).
A large part of the one-loop corrections is due to  initial-state
real and virtual photonic corrections. 
These initial-state corrections contain large logarithms of the
type $\log(s/m_e^2)$ from collinear radiation.
Such terms are universal; they are conventionally taken into account
already by convoluting the Born cross section with a radiator function,  
\beq
 \sigma(s) = \int_0^{k_{max}} {\rm d} k \, H(k) \, \sigma_0((1-k)s) \, .
\eeq
We therefore drop the universal terms from initial-state radiation (ISR)
in our one-loop result and display only the non-universal residual
corrections in the numerical presentation.

\begin{figure}[htbp]
\vspace*{1cm}
  \begin{center}
    \mbox{}\nolinebreak\hspace{15mm}
    \begin{minipage}{7.5cm}
      \includegraphics{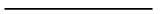}\quad {\small $M_{\rm SUSY}
        = 200 \GeV$}\hfill\\ 
      \includegraphics{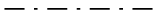}\quad {\small $M_{\rm
          SUSY} = 500 \GeV$}
    \end{minipage}\hspace{3mm}\nolinebreak
    \begin{minipage}{7.5cm}
      \includegraphics{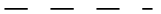}\quad {\small $M_{\rm
          SUSY} = 1000 \GeV$}\hfill\\ 
      \includegraphics{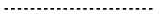}\quad {\small $M_{\rm
          SUSY} = 2000 \GeV$}
    \end{minipage}\hfill\\[0.5cm]
    \includegraphics{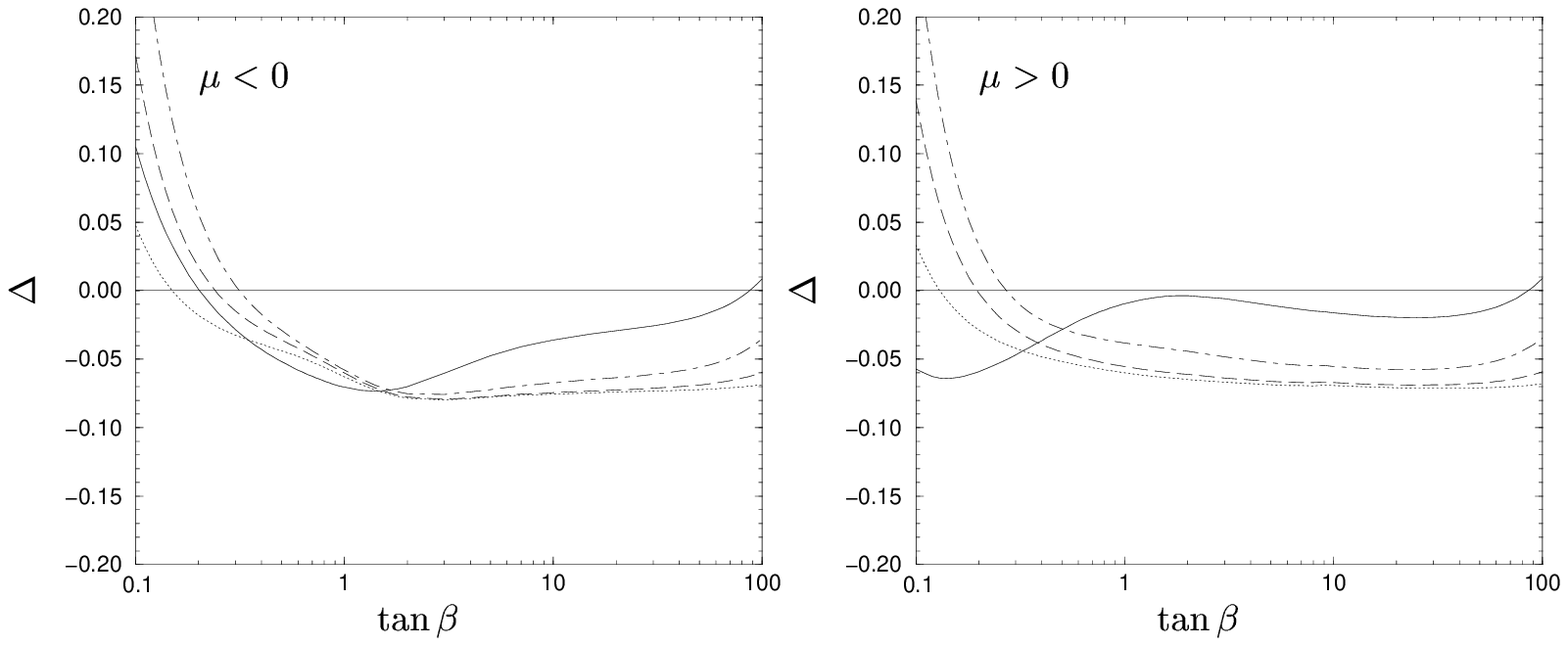}
    \mycaption{\sf The one-loop corrections to the production cross
      section of a pair of light charginos with 
      $\mch{1} = 150 \GeV$ and $M_2 = 200 \GeV$, 
      normalized to the Born approximation,
      for a center-of-mass energy of $\sqrt{s} = 500 \GeV$.}
  \label{xsectotal}
  \end{center}
\end{figure}

\bigskip
Figure~\ref{xsectotal} displays the relative one-loop correction
$\Delta$, eq.~(\ref{delta}), 
for  the production 
cross section of a light-chargino pair as a function of $\tan\beta$.
The low-$\tan\beta$ region, although experimentally ruled out 
meanwhile from Higgs-boson searches \cite{PIK},
is kept for illustration to outline the effect of the 
top-Yukawa-coupling enhancement.
The gaugino mass $M_2$ is kept fixed; varying $\tan\beta$
thus corresponds to a variation of $\mu$, which can be both
negative and positive. Typically, the higher-order contributions
are between 5 and 10\%.

%

\begin{figure}[htbp]
\vspace*{0.2cm}
  \begin{center}
    \mbox{}\nolinebreak\hspace{10mm}
    \begin{minipage}{7.5cm}
      \includegraphics{dotdashed}\quad {\small
        (s)top/(s)bottom loops}\hfill\\ 
      \includegraphics{dashed}\quad {\small
        (s)fermion loops}
    \end{minipage}\hspace{3mm}\nolinebreak
    \begin{minipage}{7.5cm}
      \includegraphics{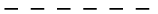}\quad {\small
        SM fermion loops}\hfill\\ 
      \includegraphics{dotted}\quad {\small complete without
        ISR}
    \end{minipage}\hfill\\[0.5cm]
    \includegraphics{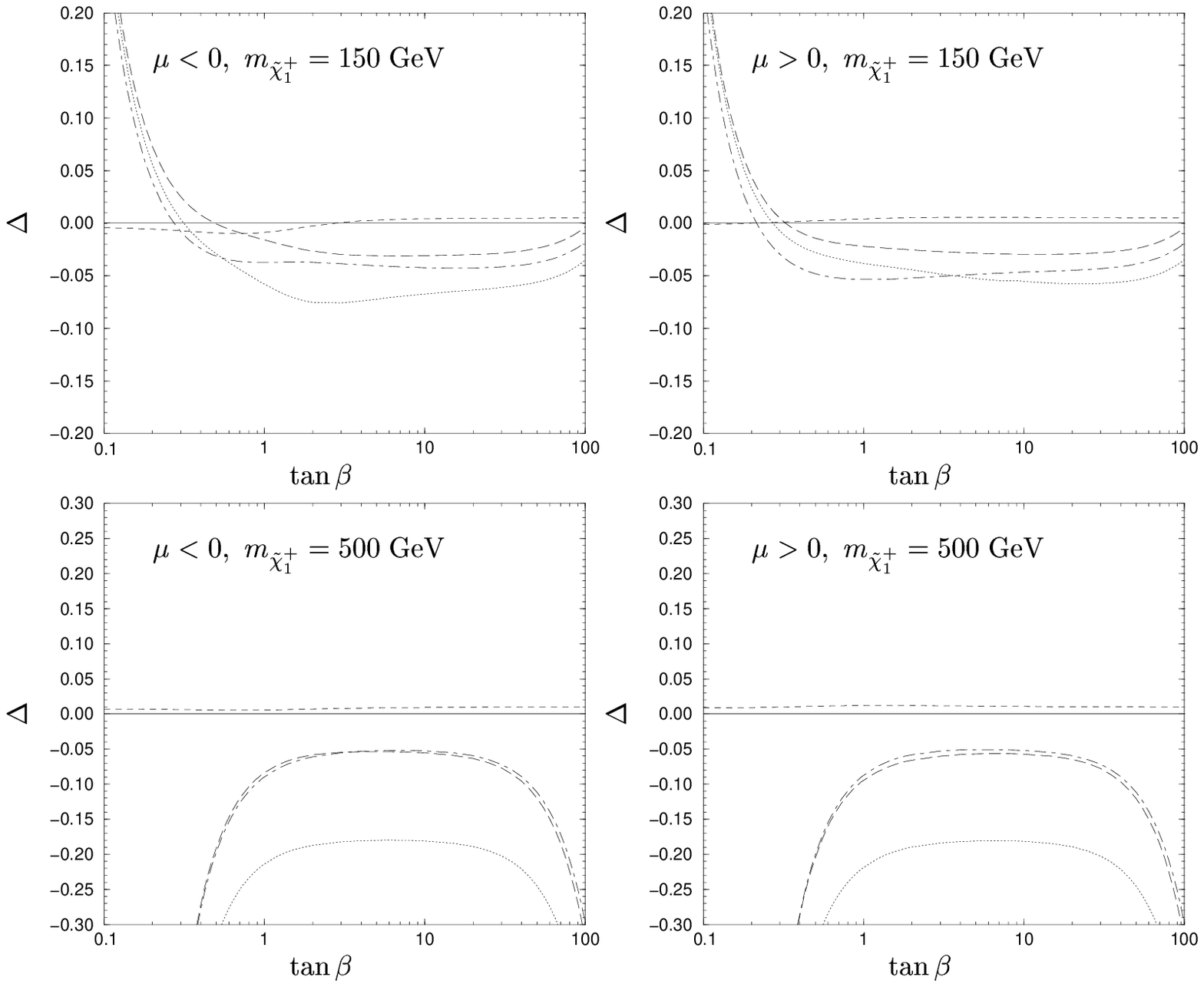}
    \mycaption{\sf One-loop corrections to the 
      chargino-pair production cross section, 
      normalized to the Born approximation.
      The various lines show the contributions from different
      subsets of diagrams. 
      Upper (lower) row: $M_2 = 200 \GeV$
      ($800 \GeV$), $\sqrt{s} = 500 \GeV$ ($1 \TeV$), 
      $M_{\rm SUSY} = 500$ GeV. }
    \label{xsecbreakdown}  
    \end{center}
\end{figure}

The breakdown of the complete set of radiative corrections 
into the various subsets of classes of diagrams is shown in
Figure~\ref{xsecbreakdown}, for two values of the lighter-chargino
mass: the subset with only top/stop and bottom/sbottom loops,
the subset with all fermion and sfermion loops, and the 
set with the complete one-loop terms. 
For illustration, also the effect resulting from
the standard-model light fermions is included. 
As expected, the standard-fermion loop contribution is very small
since its main effect is already incorporated in the Born cross 
section in terms of the running fine-structure constant.  
The figure demonstrates the comparable size of the various subclasses
and underlines the importance of a full calculation.

\begin{figure}[htbp]
\vspace{1cm}
  \begin{center}
    \includegraphics{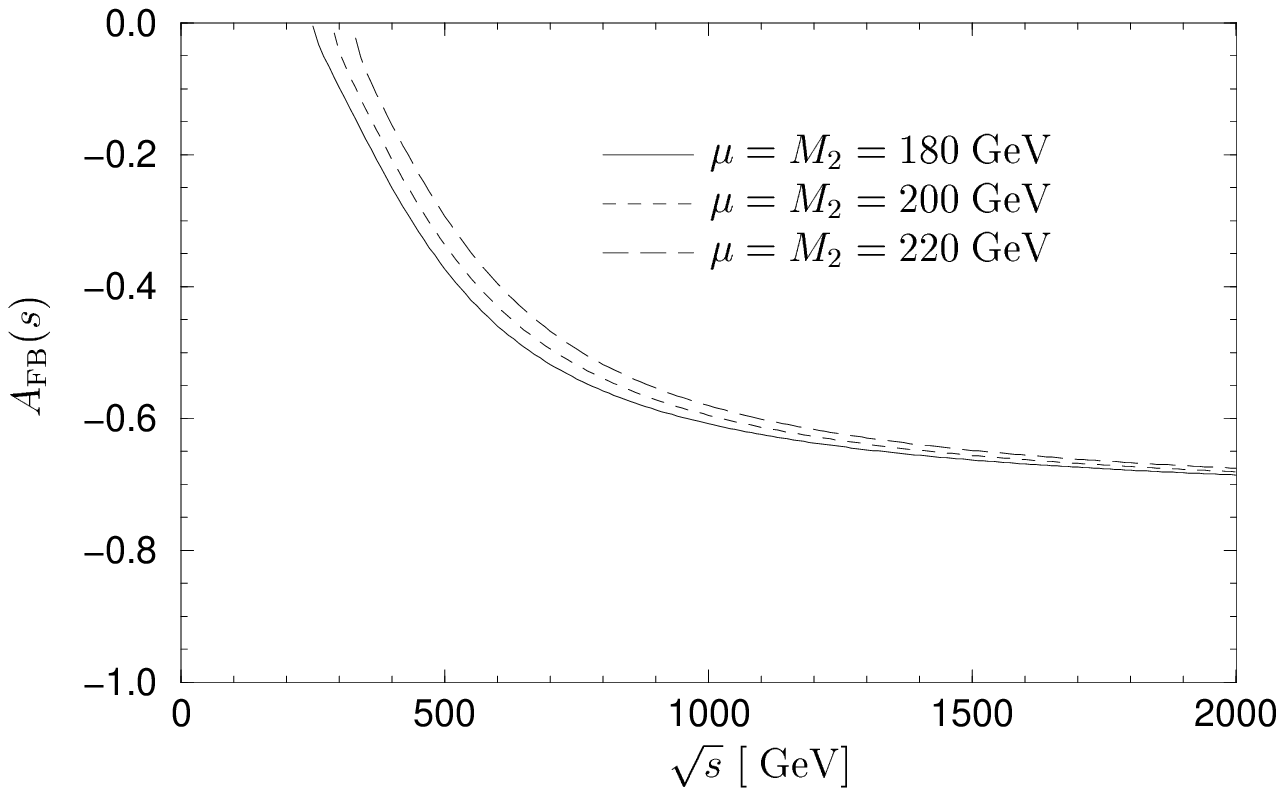}
    \mycaption{\sf The forward-backward asymmetry in Born approximation
    for the production of two light charginos
    $\epm \rightarrow \chplus_1 \chminus_1$. 
    $\tan\beta = 10$, $M_{\rm SUSY} = 300 \GeV$.}
   \label{afbborn}
  \end{center}
\end{figure}

\bigskip \noindent
(ii) {\it Forward-backward asymmetry:}
The angular distribution of the produced $\chplus_i$ shows an asymmetry,
originating at the tree level
from the different left- and right-handed couplings
to the $Z$ and from the $t$-channel-exchange diagram.
This angular asymmetry can be expressed in terms of the 
forward-backward asymmetry $A_{\rm FB}$, eq.~(\ref{afb}), which is
displayed in Figure~\ref{afbborn} as a function of the CMS energy,
for unpolarized beams.

\begin{figure}[htbp]
  \begin{center}
    \mbox{}\nolinebreak\hspace{15mm}
    \begin{minipage}{7.5cm}
      \includegraphics{solid}\quad {\small $M_{\rm SUSY}
        = 200 \GeV$}\hfill\\ 
      \includegraphics{dotdashed}\quad {\small $M_{\rm
          SUSY} = 500 \GeV$}
    \end{minipage}\hspace{3mm}\nolinebreak
    \begin{minipage}{7.5cm}
      \includegraphics{dashed}\quad {\small $M_{\rm
          SUSY} = 1000 \GeV$}\hfill\\ 
      \includegraphics{dotted}\quad {\small $M_{\rm
          SUSY} = 2000 \GeV$}
    \end{minipage}\hfill\\[0.5cm]
    \includegraphics{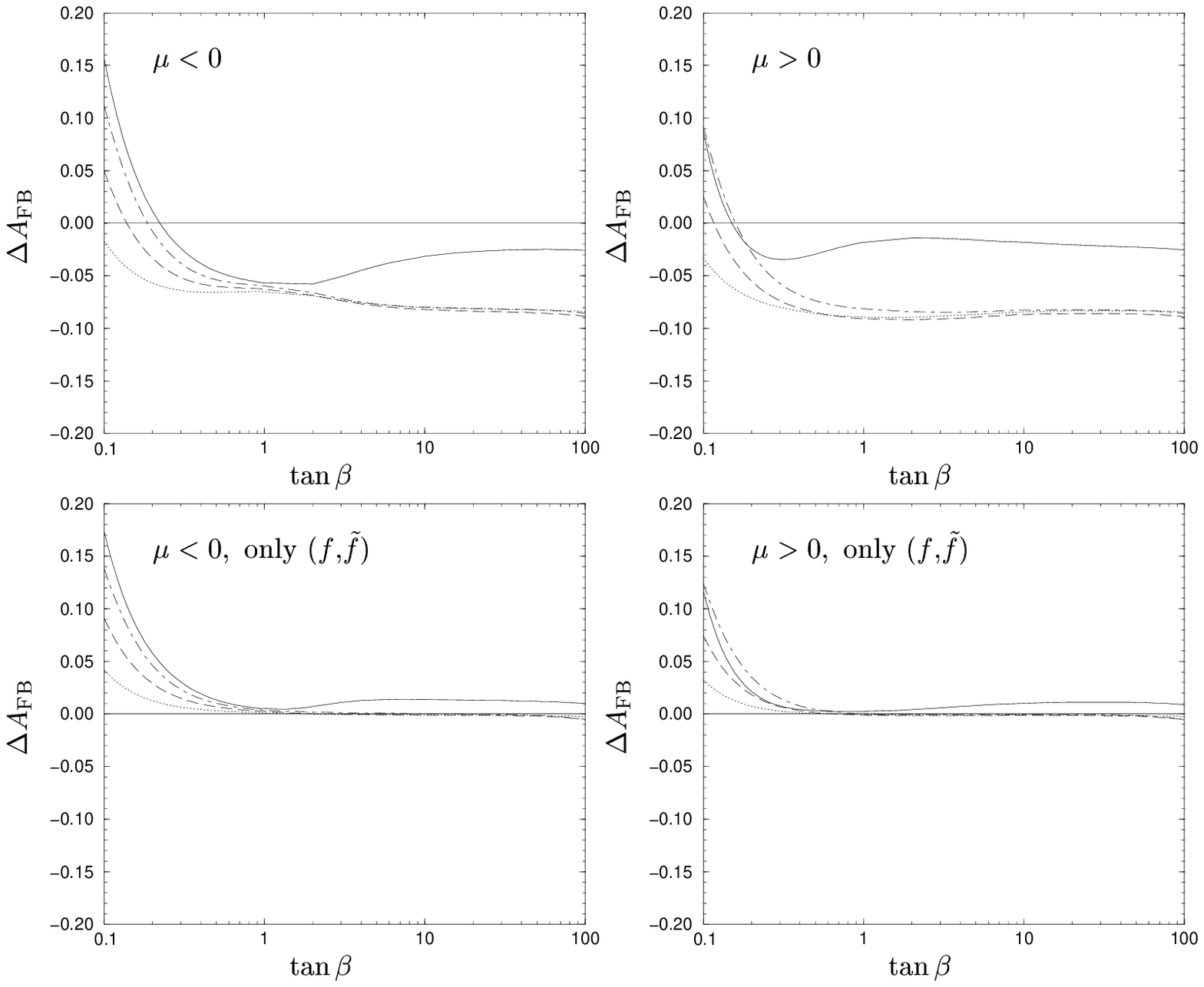}
    \mycaption{\sf The radiative corrections to the forward backward
      asymmetry for the production of two light charginos with
      $\mch{1} = 150 \GeV$, $M_2 = 200 \GeV$ and $\sqrt{s} = 500
      \GeV$. At the top, the complete corrections are shown. At the
      bottom, only the (s)fermion loops are taken into account.}
  \label{afbplot}
  \end{center}
\end{figure}

\bigskip
The loop diagrams represent an additional source for $\afb$, which
can be described in terms of a shift $\Delta\afb$ according to
\beq
 \afb = \afb^{\rm Born} + \Delta\afb \, .
\eeq
This shift is shown in Figure~\ref{afbplot} for the same set
of parameters as in Figure~\ref{xsectotal}. The loop contributions 
can thus modify the asymmetry by up to 10\%.
In order to
illustrate the total effect from the loops, 
we include 
also the contribution from the sublass of only fermion/sfermion
diagrams (lower two figures). 
The significance of the 
non-(s)fermionic loop contributions is clearly visible.

%
\clearpage

\begin{figure}[htbp]
  \begin{center}
    \mbox{}\nolinebreak\hspace{15mm}
    \begin{minipage}{7.5cm}
      \includegraphics{solid}\quad {\small $M_{\rm SUSY}
        = 200 \GeV$}\hfill\\ 
      \includegraphics{dotdashed}\quad {\small $M_{\rm
          SUSY} = 500 \GeV$}
    \end{minipage}\hspace{3mm}\nolinebreak
    \begin{minipage}{7.5cm}
      \includegraphics{dashed}\quad {\small $M_{\rm
          SUSY} = 1000 \GeV$}\hfill\\ 
      \includegraphics{dotted}\quad {\small $M_{\rm
          SUSY} = 2000 \GeV$}
    \end{minipage}\hfill\\[0.5cm]
    \includegraphics{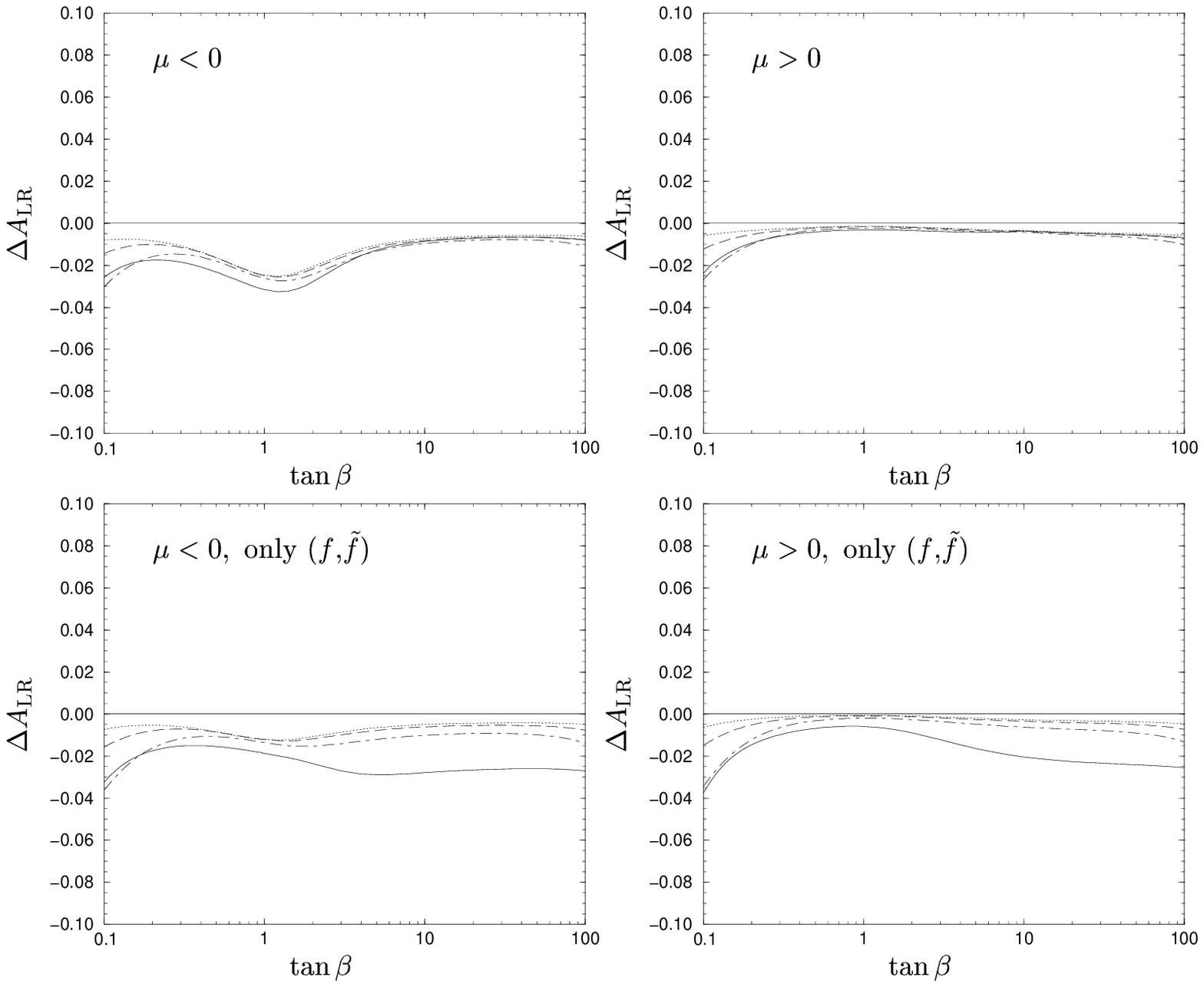}
    \mycaption{\sf The radiative corrections to the left right
      asymmetry for the production of two light charginos with
      $\mch{1} = 150 \GeV$, $M_2 = 200 \GeV$ and $\sqrt{s} = 500
      \GeV$. At the top, the complete corrections are shown. At the
      bottom, only the (s)fermion loops are taken into account.}
  \label{alrplot}
  \end{center}
\end{figure}

\noindent
(iii) {\it Left-right asymmetry:}
The left-right asymmetry $\alr$, eq.~(\ref{alr}), is of specific interest
for probing the fundamental parameters of the MSSM \cite{reconstruct},
since the $\snu$-exchange diagram is absent for right-handed electrons.
In order to point out the contributions from the higher-order terms,
we define a loop-induced shift $\Delta\alr$ with respect to the Born term,
according to
\beq
    \alr = \alr^{\rm Born} + \Delta\alr \, .
\eeq
This shift is displayed in Figure~\ref{alrplot}. Again, the same set of 
parameters has been used as in Figures~\ref{xsectotal} and~\ref{afbplot}.
For illustration, the approximate results from only (s)fermion-loop
diagrams are also shown (lower two figures).
The differences are less pronounced than in the case of $\afb$, 
but still they are significant and recommend the use of a full
calculation for precise predictions. 

%

\section{Conclusions}
A complete one-loop calculation for chargino-pair production in
electron-positron collisions has been performed, using the 
on-shell renormalization scheme. Applications to the integrated
cross sections and to the forward-backward asymmetry
$\afb$ for unpolarized beams, 
as well as to the left-right asymmetry $\alr$, have shown that 
the higher-order effects have a sizeable influence on the
theoretical predictions and therefore should be properly taken
into account for detailed studies in the MSSM.  \\[1cm]
This work has been supported by the DFG Forschergruppe 
``Quantenfeldtheorie, Computeralgebra und Monte Carlo Simulation''
and by the European Union under contract HPRN-CT-2000-00149.


\end{document}